# Cooling Dynamics of a Gold Nanoparticle in a Host Medium Under Ultrafast Laser Pulse Excitation: A Ballistic-Diffusive Approach

**M. Rashidi-Huyeh**[a, b]**, S. Volz**[a] **and B. Palpant**[b]

[a] Laboratoire d'Energétique Moléculaire et Macroscopique, Combustion (EM2C), CNRS – Ecole Centrale de Paris, Grande Voie des Vignes, 92295 Châtenay-Malabry, France

[b] Institut des Nano-Sciences de Paris (INSP), CNRS – Université Pierre et Marie Curie - Paris 6, 140 rue de Lourmel, 75015 Paris, France

## *Abstract*

We present a numerical model allowing to determine the electron and lattice temperature dynamics in a gold nanoparticle under subpicosecond pulsed excitation, as well as that of the surrounding medium. For this, we have used the electron-phonon coupling equation in the particle with a source term linked with the laser pulse, and the ballistic-diffusive equations for heat conduction in the host medium.

Our results show that the heat transfer rate from the particle to the matrix is significantly smaller than the prediction of Fourier's law. Consequently, the particle temperature rise is much larger and its cooling dynamics is much slower than that obtained using Fourier's law, which is attributed to the nonlocal and nonequilibrium heat conduction in the vicinity of the nanoparticle. These results are expected to be of great importance for interpreting pump-probe experiments performed on single nanoparticles or nanocomposite media.

## *I. Introduction*

There has been recently interest in investigating the role played by effects of thermal origin in the linear and nonlinear optical properties of noble metal nanoparticles[1]. Indeed, nanocomposite materials consisting of such particles embedded in a transparent medium exhibit specific optical response linked with the surface plasmon resonance, which is sensitive to electron distribution, electron-phonon interaction, heat transfer to the environment, and heat exchange between neighboring nano-objects. Hence, on one hand, the amplitude and temporal profile of the electromagnetic excitation determine the temperature dynamics of the particle and its environment[2]. On the other hand, this thermal response modifies the material optical properties themselves through mechanisms of different nature (hot electrons, thermal lensing,…)[2-5]. While noble metal nanoparticles have been proposed for many applications such as photonic (*plasmonic*) devices, chemical- or bio-sensing, medical imaging or even thermal therapy[6-8], the fundamental study of the physical mechanisms governing the coupling between light, electrons and phonons has become a challenging issue. Pump-probe time-resolved spectroscopy represents certainly one of the most efficient tools for performing such investigations from the experimental point of view. Thermal phenomena of different origins can thus be generated and their dynamics further studied by using ultrafast pulsed lasers. In this case, Fourier's law is no longer suited to describe heat propagation as the spatial and time scales under consideration are smaller than the heat carrier mean free path and lifetime, respectively[9–13].

In a previous work[2], we have presented a theoretical approach to determine the temperature dynamics of nanocomposite materials consisting of metal nanoparticles dispersed in a host dielectric medium under pulsed laser excitation (improved three temperatures model). For this, the usual electron-phonon coupling, the particle-matrix thermal transfer at the interface and the heat diffusion (Fourier's law) in the matrix have been considered. However, it has been shown that the heat transfer cannot be correctly explained by the Fourier approach when the medium characteristic scale (and/or the characteristic time of the heat variation rate) becomes as small as the heat carrier mean free path (and/or the heat carrier lifetime). In these cases, the Boltzmann transport equation (BTE) is often proposed to analyze the heat transfer in the medium[9]. An alternative method, named the ballistic-diffusive equation (BDE), based on the BTE in the framework of the relaxation time approximation, has been suggested by Chen. This approach consists in splitting the heat flux intensity and hence the internal energy at any point of the matrix into two components: one represents ballistic processes originating from the boundaries and the other corresponds to diffusive processes[10–13].

In this paper we present a theoretical approach to determine the electron and lattice temperature dynamics of a metal nanoparticle (gold here) embedded in a dielectric host medium with finite volume, as well as the internal energy (or local temperature) dynamics at each point of this host medium. We then apply the BDE to evaluate the heat transfer from the metal nanoparticle and the heat propagation through the surrounding dielectric matrix.

## II. Model

We consider a metallic spherical nanoparticle with radius $R_p$ surrounded by a dielectric shell with internal and external radiuses $R_p$ and $R_e$, respectively. The shell thickness is then $d_s = R_e - R_p$. Let us mention that such core-shell shaped nanoparticles have already been synthesized[14] and investigated for their linear and nonlinear optical properties[15,16]. The case of an isolated nanoparticle embedded in a host medium, or the one of a diluted nanocomposite material, corresponds to the infinite shell thickness limit and can also be accounted for by the present model.

As we have already described in Ref. 2, a light pulse is partly absorbed by the conduction electron gas of the metal particle. This energy is transferred to the particle lattice via electron-phonon coupling and then relaxes into the dielectric shell. We consider the same assumptions, initial and boundary conditions as in this previous work. The evolution of the electron temperature, $T_e$, is then driven by:

$$C_e \frac{\partial T_e}{\partial t} = -G(T_e - T_l) + P_{vol}(t), \qquad (1)$$

where $C_e = \gamma_e T_e$ is the electron heat capacity, $\gamma_e$ is a constant, $G$ is the electron-phonon coupling constant, $P_{vol}(t)$ denotes the instantaneous power absorbed per metal volume unit and $T_l$ is the lattice temperature. For the temporal evolution of the lattice temperature we have:

$$V_p C_l \frac{\partial T_l}{\partial t} = V_p G(T_e - T_l) - H(t), \qquad (2)$$

where $C_l$ is the heat capacity of the lattice, $V_p$ is the volume of the particle and $H(t)$ is the instantaneous heat power transferred to the dielectric shell from the particle. $H(t)$ may be written as a function of the heat flux density, $\mathbf{q}(\mathbf{r},t)$, integrated over the particle surface, $S_p$:

$$H(t) = \int_{S_p} \mathbf{q}(\mathbf{r},t) \cdot \mathbf{n}\, ds, \qquad (3)$$

where **n** is the outward unit vector normal to the particle surface. In the aim to determine $\mathbf{q}(\mathbf{r},t)$, we have applied the ballistic-diffusive approximation[12]. The essence of this approximation stands on the splitting of the carrier heat distribution function (and so the local internal energy and heat flux) at any point of the matrix into two components: one is due to carriers coming directly from the boundaries (particle surface), without any scattering, and represents the ballistic component. The other is related to the carriers which arrive after scattering from other points of the matrix, and may be approximated as a diffusive process. Details of the computation will be presented in a full-length paper. We only report here the main results, which will be expressed in term of the following nondimensional parameters:

- nondimensional time $t^* = t/\tau$, where $\tau$ is the phonon relaxation time in the dielectric medium,

- nondimensional radial distance $r^* = r/\Lambda_m$, where $\Lambda_m$ denotes the mean free path of the heat carriers in the dielectric which may be given by $\sqrt{3D_m\tau}$ with $D_m$ the matrix heat diffusion constant,

- electron nondimensional temperature $\Theta_e(t) = \dfrac{T_e(t) - T_0}{T_0}$, where $T_0$ is the ambient temperature,

- lattice nondimensional temperature $\Theta_l(t) = \dfrac{T_l(t) - T_0}{T_0}$ .

## III. Results and discussion

We have applied this model to a gold nanoparticle core and an alumina (amorphous $Al_2O_3$) shell with thermodynamic properties corresponding to their bulk phase[2]: $G = 3 \times 10^{16}\,\text{W m}^{-3}\,\text{K}^{-1}$, $\gamma = 66\,\text{J m}^{-3}\,\text{K}^{-2}$, $C_l = 2.49 \times 10^6\,\text{J m}^{-3}\,\text{K}^{-1}$, $D_m = 1.16 \times 10^{-5}\,\text{m}^2\,\text{s}^{-1}$, $\tau = 0.85\,\text{ps}$ and $\Lambda_m = 5.4\,\text{nm}$. The initial temperature, $T_0$, is fixed at 300 K. Moreover, $P_{vol}(t)$ is considered to exhibit the same time dependence as the incident pulse, which is supposed to be a Gaussian. It is also to mention that as the spatial width of the pulse (few tenths of micrometers) is much larger than the particle size (from a few nanometers to tenths of nanometers), the instantaneous energy absorbed by the particle is homogenous over the particle volume. We have then: $P_{vol}(t) = Ae^{-B(t-t_0)^2}$ where the parameter values are chosen equal to those of our earlier work[2] (*i.e.* $A = 1.4 \times 10^{21}$ W m$^{-3}$, $B = 2.3 \times 10^{26}$ s$^{-2}$ and $t_0 = 150$ fs). The corresponding pulse duration is 110 fs.

Let us first examine the results of this model, based on the BDE, and those based on Fourier's law. For this, we have calculated the temperature dynamics using these two approaches with the same values and conditions. Figure 1 presents, with a vertical logarithmic scale, the electron and lattice nondimensional temperature dynamics (black and grey lines) obtained using BDE (solid line) and Fourier one (dash line). The gold particle radius is 10 nm and the shell thickness, $d_s$, is of the order of $\Lambda_m$ ($\approx 5.4$ nm). As we can see, these two methods predict the same behaviour for the electron temperature during a few initial picoseconds, where it increases very rapidly up to 2200 K ($\Theta_e = 6.3$) just after the pulse passage. It then presents a rapid relaxation due to electron-phonon scattering. As it can be seen on the figure, $T_l$ is very low as compared to $T_e$ during this time and can then be neglected in equation 1. On the other hand the pulse is quickly off during at the beginning of this short time regime. One can then easily show that the electron temperature may be simply obtained as:

$$T_e(t) - T_{e,\max} \simeq -\frac{G}{\gamma_e} t$$

where $T_{e,\max}$ is the maximum value of the electron gas temperature. The characteristic time of the rapid relaxation $\tau_r$, defined as the time necessary for the electron temperature to reach the half of its maximum, is $\tau_r = \frac{T_{e,\max} \gamma_e}{2G}$. For the values considered here, this time is about $5\tau \simeq 4$ ps. At the end of this relaxation, the electrons and lattice attain a thermal equilibrium and we can indifferently denote their respective temperatures by the term *particle temperature*. Whatever the theoretical approach, this thermal equilibrium occurs after $\sim 15\tau$.

This rapid relaxation is then followed by a slow one linked with the heat transfer to the matrix. This is the reason for which it depends crucially on the heat transfer mechanism: while the Fourier law predicts a rapid decay, the BDE presents a much more slow one. In fact, the Fourier theory is valid only when there are enough scattering events in the matrix such that carriers can exchange energy with the matrix. This assumption leads to an over-prediction of the heat release from the particle and is, for this reason, to be evidently avoided in the nanoscale and/or the subpicosecond heating process cases. Therefore, the heat transfer from the particle is actually slower than envisaged by the Fourier theory. This then leads to a characteristic time for the slow relaxation more elevated in the BDE case than in the Fourier one. In addition, as the energy injected by the pulse stays for a longer time in the particle, the lattice temperature predicted by the BDE is higher and its maximum is reached later than with the calculation using Fourier's law (Fig. 1).

As the shell thickness plays an important role in the heat transfer mechanism, let us investigate now its influence on the particle temperature relaxation. Figure 2 presents the electron temperature dynamics for different values of the shell thickness, $d_s$, from $0.1\Lambda_m = 0.54$ nm to $10\Lambda_m = 54$ nm. As can be seen on Fig. 2, $\Theta_e(t)$ is independent of $d_s$ during the short time regime (rapid relaxation). But after a few picoseconds, $\Theta_e(t)$ evolves depending on this parameter: the thicker the shell, the larger the characteristic relaxation time.

These behaviours may be explained as follows: the electron-phonon characteristic time being worth a few picoseconds, the energy transfer to the matrix is not effective during the short time regime. Consequently, $\Theta_e(t)$ is not sensitive to the morphology of the particle environment and the discussion given above regarding the characteristic time of the rapid relaxation remains valid. Now when the heat transfer to the matrix becomes effective, the shell thickness value can affect the particle temperature. Indeed, for a shell the thickness of which is inferior to or comparable with $\Lambda_m$, the heat transfer mainly occurs as a ballistic mechanism, the diffusive one remaining weak. When $d_s$ increases the contribution of diffusive processes to the heat transfer becomes more important, and as we can see on Fig. 2 the slow relaxation characteristic time tends to a certain constant value. This is the "diffusive limit" which is reached when the shell thickness exceeds a few $\Lambda_m$. For clarifying this, we have compared on Fig. 3 the time dependence of $\Theta_e$ obtained using the BDE and Fourier approaches for a shell thickness equivalent to $\Lambda_m$ and $10\Lambda_m$. As we can see on this figure the slow relaxation time predicted by these two approaches is quite different for $d_s = \Lambda_m$, while for $d_s = 10\Lambda_m$ it is the same. However, the particle temperature obtained using the BDE remains one order of magnitude higher than that provided by the Fourier law. This can be ascribed to the nonlocal heat transfer around the particle[10].

We would finally like to underline two points regarding the model. The first one concerns the difference between the boundary conditions in the BDE model and the Fourier one. In fact, the values of the temperatures used in the boundary conditions for the BDE are those of the emitted phonons (non-reflecting condition), while in the Fourier approach they are those of equilibrium phonons. This causes an artificial temperature jump at the interface in the BDE case. The second point is related to the limit of the ballistic-diffusive approach. As the diffusive contribution in the BDE is only an approximation to the heat carrier scattering processes, this model is less accurate in the steady state and when the diffusive component is dominant[12,13].

In summary, we have presented here a model allowing to determine the electron and lattice temperature dynamics in a gold nanoparticle, embedded in a finite volume dielectric host medium, under a subpicosecond laser pulse. This model is based on the usual electron-phonon coupling in the particle and the ballistic-diffusive approximation in the dielectric medium. It is well suited to such situation where the Fourier theory is disqualified to describe the heat propagation correctly at the space and time scales involved. For a thin dielectric shell, the prediction of the Fourier law for the slow relaxation characteristic time deviates from that predicted using the BDE. As the shell thickness increases we tend to a "diffusive limit" for which the slow relaxation time is comparable with that obtained using the Fourier approach for heat conduction. However, the lattice temperature determined through the BDE is one order of magnitude higher than with Fourier's law. Confirming these results by ultrafast pump-probe experiments now represents a challenging issue.

**Acknowledgements**. This work has been supported by the program PNANO of the Agence Nationale de la Recherche and by the region Ile-de-France, under the project SESAME E-1751.

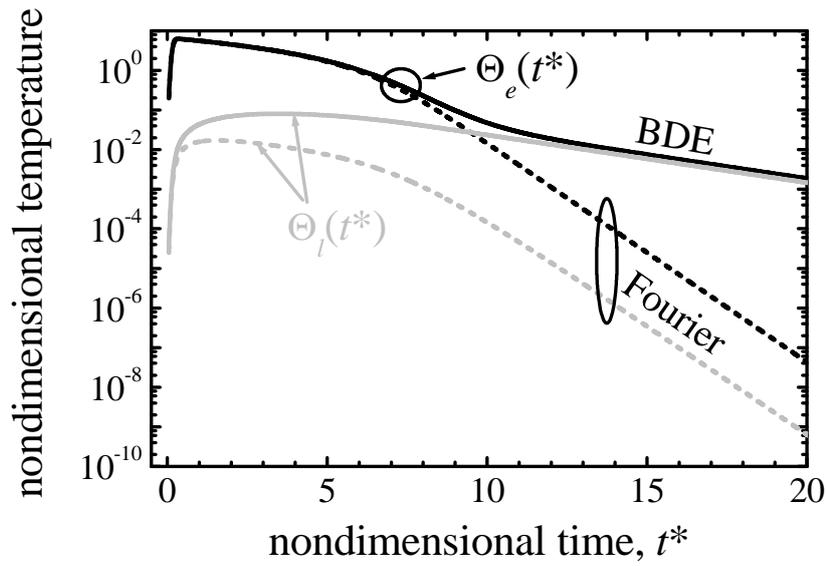

Figure 1: Nondimensional electron and lattice temperature dynamics [$\Theta_e(t^*)$ and $\Theta_l(t^*)$, respectively] of a gold nanoparticle surrounded by an alumina shell under an ultrashort laser pulse excitation, obtained using the ballistic-diffusive approximation (solid lines) and Fourier's law for the heat conduction (dash lines). The nanoparticle radius is 10 nm and the shell thickness is equivalent to the phonon mean free path $\Lambda_m \approx 5.4$ nm. The laser pulse duration is 110 fs.

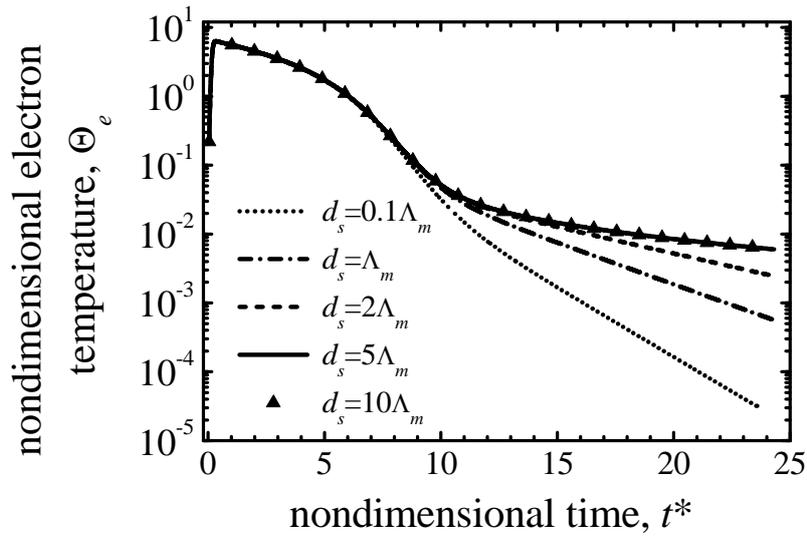

Figure 2: Nondimensional electron temperature dynamics of a gold nanoparticle (10 nm radius) in an alumina shell the thickness of which, $d_s$, varies from $0.1\Lambda_m \square 0.54$ nm to $10\Lambda_m \square 54$ nm, calculated by solving the BDE.

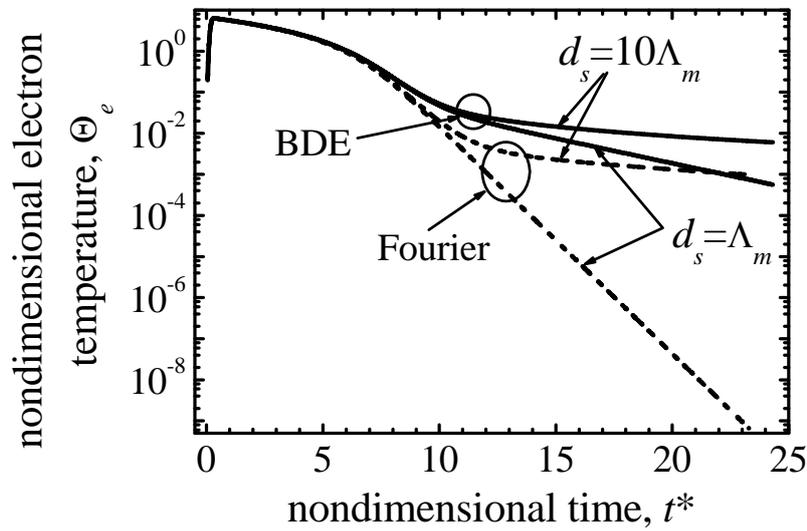

Figure 3: Nondimensional electron temperature dynamics of a gold nanoparticle (10 nm radius) in an alumina shell with two different thickness values, obtained using the BDE (solid lines) and Fourier's law (dash lines).

# References

[1] B. Palpant, in "Nonlinear optical properties of matter: From molecules to condensed phases", Series: Challenges and Advances in Computational Chemistry and Physics, Vol. 1, pp. 461–508, edited by M. G. Papadopoulos, J. Leszczynski and A. J. Sadlej (Springer, 2006).

[2] M. Rashidi-Huyeh and B. Palpant, J. Appl. Phys. **96**, 4475 (2004).

[3] M. Rashidi-Huyeh and B. Palpant, Phys. Rev. B **74**, 075405 (2006).

[4] B. Palpant, M. Rashidi-Huyeh, B. Gallas, S. Chenot, and S. Fisson, Appl. Phys. Lett. **90**, 223105 (2007).

[5] B. Palpant, D. Prot, A.-S. Mouketou-Missono, M. Rashidi-Huyeh, C. Sella, and S. Debrus, Proc. of SPIE **5221**, 14 (2003).

[6] S. Eustis and M. A. El-Sayed, Chem. Soc. Rev. **35**, 209 (2006).

[7] S. Berciaud, L. Cognet, G. A. Blab, and B. Lounis, Phys. Rev. Lett. **93**, 257402 (2004).

[8] D. P. O'Neal, L. R. Hirsch, N. J. Halas, J. D. Payne, and J. L. West, Cancer Lett. **209**, 171 (2004).

[9] A. Majumdar, J. Heat Transfer **115**, 327 (1993).

[10] G. Chen, J. Heat Transfer ASME **118**, 539 (1996).

[11] G. Chen, Phys. Rev. Lett. **86**, 2297 (2001).

[12] G. Chen, J. Heat Transfer ASME **124**, 320 (2002).

[13] R. Yang, G. Chen, M. Laroche, and Y. Taur, J. Heat Transfer ASME **127**, 298 (2005).

[14] J. J. Penninkhof, T. van Dillen, S. Roorda, C. Graf, A. van Blaaderen, A. M. Vredenberg, and A. Polman, Nucl. Instr. and Meth. in Phys. Res. B **242**, 523 (2006).

[15] A. V. Goncharenko, Chem. Phy. Lett. **386**, 25 (2004).

[16] A. V. Goncharenko and Y.-C. Chang, Chem. Phy. Lett. **439**, 121 (2007).